\documentclass[conference]{IEEEtran}
\IEEEoverridecommandlockouts
\usepackage{tikz}
\usepackage{pgfplots}
\pgfplotsset{compat=newest}
\pgfplotsset{plot coordinates/math parser=false}
\newlength\figureheight
\newlength\figurewidth
\usetikzlibrary{plotmarks}
\usetikzlibrary{arrows.meta}
\usepackage{cite}
\usepackage{amsmath,amssymb,amsfonts}
\usepackage{algorithmic}
\usepackage{graphicx}
\usepackage{textcomp}
\usepackage{xcolor}
\usepackage{units}
\usepackage{multirow}
\usepackage{booktabs}
\usepackage{caption}
\usepackage{subcaption}

\usepackage[format=hang]{caption}       
\def\BibTeX{{\rm B\kern-.05em{\sc i\kern-.025em b}\kern-.08em
    T\kern-.1667em\lower.7ex\hbox{E}\kern-.125emX}}
\begin{document}

\title{Radar-based Respiratory Rate Monitoring in Standing Position\\

}

\author{\IEEEauthorblockN{Tassneem Helal \textsuperscript{1}\IEEEauthorrefmark{1}, Fady Aziz\textsuperscript{1}\IEEEauthorrefmark{1}, Omar Metwally\textsuperscript{1},  Marco F.~Huber\textsuperscript{2}, \\ 
		Dominik Alscher\textsuperscript{3}, Christoph Wasser\textsuperscript{3}, Urs Schneider\textsuperscript{1}
		\thanks{This study was realized supported by a feasibility grant by Iturri S.A., and was co-funded by BMVG under project reference E/U2ED/LD007/KF557.}
		\thanks{Email: fady.aziz@ipa.fraunhofer.de}}
		\thanks{\textbf{*The first two authors contributed equally to this work.}}

	\IEEEauthorblockA{\textsuperscript{~1}Department of Bio-mechatronic Systems, Fraunhofer IPA, Stuttgart, Germany}
	\IEEEauthorblockA{\textsuperscript{~2}Center for Cyber Cognitive Intelligence, Fraunhofer IPA, Stuttgart, Germany}
	\IEEEauthorblockA{\textsuperscript{~3}Robert-Bosch-Hospital, Stuttgart, Germany}
	\vspace{-0.3 cm}
}

\maketitle

\begin{abstract}
\par Estimating human vital signs in a contactless non-invasive method using radar provides a convenient method in the medical field to conduct several health checkups easily and quickly. In addition to monitoring while sitting and sleeping, the standing position has aroused interest for both the industrial and medical fields. However, it is more challenging due to the micro motions induced by the body for balancing that may cause false respiratory rate estimation. In this work, we focus on the measurement of the respiratory rate of a standing person accurately with the capability of heavy breath detection and estimation. Multiple estimation approaches are presented and compared, including spectral estimation, deep-learning-based approaches, and adaptive peak selection with Kalman filtering. The latest technique is showing the best performance with an absolute error rate of \unit[$\approx\pm 1.5$]{bpm}, when compared to a Vernier Go Direct\textsuperscript{\textregistered} respiration belt.
\end{abstract}

\begin{IEEEkeywords}
Vital Signs, FMCW Radar, Health Care
\end{IEEEkeywords}

\section{Introduction}

\par An irregular respiratory rate is an essential indication of severe respiratory disorder and a crucial sign used to track the progression of sickness \cite{al2011respiration}. A series of studies over the years have shown that respiratory rate could be one of the most beneficial vital signs to track \cite{fieselmann1993respiratory}\cite{subbe2003effect}\cite{goldhill2005physiologically}\cite{cretikos2007objective}. Especially during the current Corona crisis, respiratory rate is considered to be an indication whether a patient might have the infection or not and the severity of the medical situation \cite{massaroni2020remote}. According to the WHO (World Health Organization), a resting value of respiratory rate greater than 30 can be used as a critical sign for severe pneumonia in adults \cite{world2020clinical}. Contact-based respiration monitoring, such as acoustic-based methods \cite{werthammer1983apnea}, airflow-based methods, chest and abdominal movement detection, transcutaneous CO\textsubscript{2} monitoring, ECG (Electrocardiogram), and Oximetry probe has achieved remarkable results. However, contactless methods for estimating the respiratory rate are in great need, especially during the Corona pandemic crisis. 

\par Radar has been investigated as a feasible sensor for vital signs monitoring due to its contact-less monitoring capability. Various state-of-the-art techniques are based on analyzing the rate of change of the phase of the estimated target location by the radar \cite{8378778,ahmad2018vital}. Such techniques are based on monitoring the human while being stationary while sitting or sleeping \cite{8777864}. Thus, it is ensured that the human body is not inducing any motion or micro-motion except the chest motion, which is considered as the main motion affected by the cardiac and respiration activities. The presented techniques could achieve a reliable estimation accuracy in comparison to the classic tools that are based on direct contact with the body. However, such studies will be infeasible in other stationary situations e.g., the standing position. While standing, the human body induces other micro motions for weight balancing that are expected to be estimated by the radar as a false respiratory rate. Some studies had a similar focus on body random movements mitigation \cite{6697618, 9173434}. However, such techniques rely on the fact that such motions are showing behaviors different from the vital signs. This is not the case in the standing posture, as the weight balancing motion will be realized by the radar as a similar behavior to the chest activity.

\par The main focus of this study is analyzing the standing posture for accurate estimation of respiratory rate. The study has a primary focus of adding an adaptive feature to the state-of-the-art techniques, such that the breath harmonic can be tracked accurately. The study has a direct use-case scenario, where the radar module is integrated with a thermal camera that is used for detecting the body temperature. The system is presented by the Fraunhofer institute as a prototype for Covid-19 virus detection that can be used in public places e.g., airports and hospital entrances \cite{fraunhofer_2020}. Additionally, the main focus of the study is directed towards the respiratory rate estimation, not the heart rate, due to the higher demand and interest in the medical field. 

\par The remainder of this document is structured as follows: Section ~\ref{sec:methodology} introduces the methodology of the proposed system, Section ~\ref{sec:exp} introduces the experimental setup for the respiratory rate measurements for standing humans. While Section ~\ref{sec:results}  presents the results of the accuracy analysis. Section ~\ref{sec:conc} presents the conclusion and a discussion about future work. 

\section{Methodology}\label{sec:methodology}

\par
\begin{figure}[t!]
    \centering
    \includegraphics[width=0.45\textwidth]{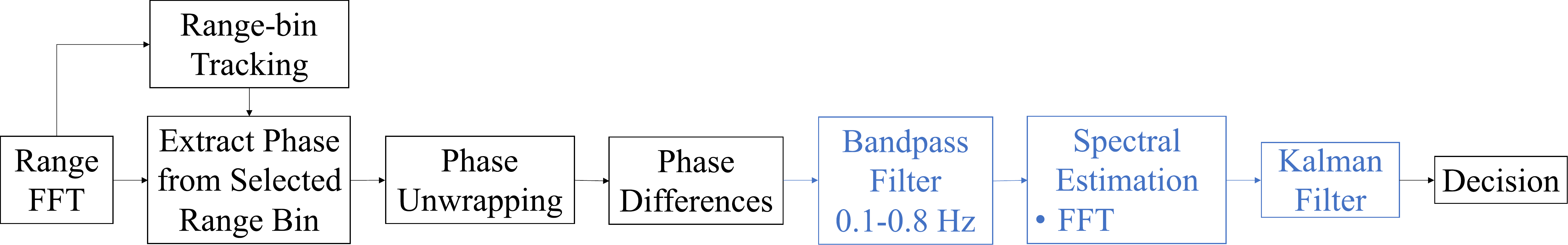}
	\caption{\centering Implemented Respiratory Rate Estimation Block Diagram}
	\label{fig:newcodebd}
\end{figure}

 Our primary goal is to implement and design a radar system that can accurately measure the vital signs of a standing target with a large spectrum of respiratory rate values on a real-time basis. The proposed signal processing chain in previous studies is applied to sitting targets to limit the effect of random body movements. However, it is more convenient for public places scenarios to measure in the standing position. The utilized bandpass filter in such techniques is limited to a bandwidth of \unit[0.1-0.5]{Hz} which maps to a respiratory rate range of \unit[6-30]{bpm}. In order to be able to detect severe cases of heavy breath occurrences, the cut-off frequency of the bandpass filter is extended in our implementation to $\unit[0.1-0.8]{Hz}$. The new bandwidth will give us a range of values of $\unit[6-50]{bpm}$. Enhancements were made on the previously mentioned signal processing chain to achieve our desired goals. Fig.~\ref{fig:newcodebd} shows the proposed signal processing chain, where the blue colored blocks represent areas of change and addition to overcome the induced body micro motions in standing position. 

\begin{figure}[t!]
    \centering
    \includegraphics[width=0.45\textwidth]{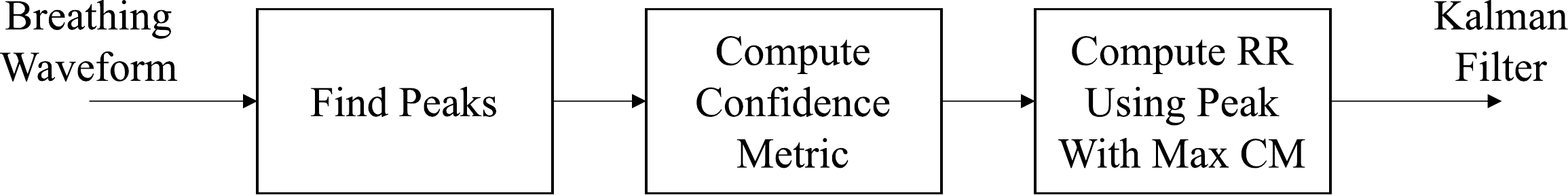}
	\caption{Spectral Estimation Algorithm}
	\label{fig:specest}
	\vspace{-0.5 cm}
\end{figure}

\par
\begin{figure}[t!]
    \centering
         {\resizebox{0.98\columnwidth}{!}{
%
%
\begin{tikzpicture}

\begin{axis}[%
width=4.521in,
height=2.5in,
at={(1.962in,0.797in)},
scale only axis,
point meta min=2000,
point meta max=26868,
axis on top,
xmin=-0.50125313283208,
xmax=400.501253132832,
xlabel style={font=\color{white!15!black}},
xlabel={\Large Frame},
ymin=0.491025641025641,
ymax=1.20897435897436,
ylabel style={font=\color{white!15!black}},
ylabel={\Large Range [m]},
axis background/.style={fill=white},
colormap={mymap}{[1pt] rgb(0pt)=(0.2422,0.1504,0.6603); rgb(1pt)=(0.2444,0.1534,0.6728); rgb(2pt)=(0.2464,0.1569,0.6847); rgb(3pt)=(0.2484,0.1607,0.6961); rgb(4pt)=(0.2503,0.1648,0.7071); rgb(5pt)=(0.2522,0.1689,0.7179); rgb(6pt)=(0.254,0.1732,0.7286); rgb(7pt)=(0.2558,0.1773,0.7393); rgb(8pt)=(0.2576,0.1814,0.7501); rgb(9pt)=(0.2594,0.1854,0.761); rgb(11pt)=(0.2628,0.1932,0.7828); rgb(12pt)=(0.2645,0.1972,0.7937); rgb(13pt)=(0.2661,0.2011,0.8043); rgb(14pt)=(0.2676,0.2052,0.8148); rgb(15pt)=(0.2691,0.2094,0.8249); rgb(16pt)=(0.2704,0.2138,0.8346); rgb(17pt)=(0.2717,0.2184,0.8439); rgb(18pt)=(0.2729,0.2231,0.8528); rgb(19pt)=(0.274,0.228,0.8612); rgb(20pt)=(0.2749,0.233,0.8692); rgb(21pt)=(0.2758,0.2382,0.8767); rgb(22pt)=(0.2766,0.2435,0.884); rgb(23pt)=(0.2774,0.2489,0.8908); rgb(24pt)=(0.2781,0.2543,0.8973); rgb(25pt)=(0.2788,0.2598,0.9035); rgb(26pt)=(0.2794,0.2653,0.9094); rgb(27pt)=(0.2798,0.2708,0.915); rgb(28pt)=(0.2802,0.2764,0.9204); rgb(29pt)=(0.2806,0.2819,0.9255); rgb(30pt)=(0.2809,0.2875,0.9305); rgb(31pt)=(0.2811,0.293,0.9352); rgb(32pt)=(0.2813,0.2985,0.9397); rgb(33pt)=(0.2814,0.304,0.9441); rgb(34pt)=(0.2814,0.3095,0.9483); rgb(35pt)=(0.2813,0.315,0.9524); rgb(36pt)=(0.2811,0.3204,0.9563); rgb(37pt)=(0.2809,0.3259,0.96); rgb(38pt)=(0.2807,0.3313,0.9636); rgb(39pt)=(0.2803,0.3367,0.967); rgb(40pt)=(0.2798,0.3421,0.9702); rgb(41pt)=(0.2791,0.3475,0.9733); rgb(42pt)=(0.2784,0.3529,0.9763); rgb(43pt)=(0.2776,0.3583,0.9791); rgb(44pt)=(0.2766,0.3638,0.9817); rgb(45pt)=(0.2754,0.3693,0.984); rgb(46pt)=(0.2741,0.3748,0.9862); rgb(47pt)=(0.2726,0.3804,0.9881); rgb(48pt)=(0.271,0.386,0.9898); rgb(49pt)=(0.2691,0.3916,0.9912); rgb(50pt)=(0.267,0.3973,0.9924); rgb(51pt)=(0.2647,0.403,0.9935); rgb(52pt)=(0.2621,0.4088,0.9946); rgb(53pt)=(0.2591,0.4145,0.9955); rgb(54pt)=(0.2556,0.4203,0.9965); rgb(55pt)=(0.2517,0.4261,0.9974); rgb(56pt)=(0.2473,0.4319,0.9983); rgb(57pt)=(0.2424,0.4378,0.9991); rgb(58pt)=(0.2369,0.4437,0.9996); rgb(59pt)=(0.2311,0.4497,0.9995); rgb(60pt)=(0.225,0.4559,0.9985); rgb(61pt)=(0.2189,0.462,0.9968); rgb(62pt)=(0.2128,0.4682,0.9948); rgb(63pt)=(0.2066,0.4743,0.9926); rgb(64pt)=(0.2006,0.4803,0.9906); rgb(65pt)=(0.195,0.4861,0.9887); rgb(66pt)=(0.1903,0.4919,0.9867); rgb(67pt)=(0.1869,0.4975,0.9844); rgb(68pt)=(0.1847,0.503,0.9819); rgb(69pt)=(0.1831,0.5084,0.9793); rgb(70pt)=(0.1818,0.5138,0.9766); rgb(71pt)=(0.1806,0.5191,0.9738); rgb(72pt)=(0.1795,0.5244,0.9709); rgb(73pt)=(0.1785,0.5296,0.9677); rgb(74pt)=(0.1778,0.5349,0.9641); rgb(75pt)=(0.1773,0.5401,0.9602); rgb(76pt)=(0.1768,0.5452,0.956); rgb(77pt)=(0.1764,0.5504,0.9516); rgb(78pt)=(0.1755,0.5554,0.9473); rgb(79pt)=(0.174,0.5605,0.9432); rgb(80pt)=(0.1716,0.5655,0.9393); rgb(81pt)=(0.1686,0.5705,0.9357); rgb(82pt)=(0.1649,0.5755,0.9323); rgb(83pt)=(0.161,0.5805,0.9289); rgb(84pt)=(0.1573,0.5854,0.9254); rgb(85pt)=(0.154,0.5902,0.9218); rgb(86pt)=(0.1513,0.595,0.9182); rgb(87pt)=(0.1492,0.5997,0.9147); rgb(88pt)=(0.1475,0.6043,0.9113); rgb(89pt)=(0.1461,0.6089,0.908); rgb(90pt)=(0.1446,0.6135,0.905); rgb(91pt)=(0.1429,0.618,0.9022); rgb(92pt)=(0.1408,0.6226,0.8998); rgb(93pt)=(0.1383,0.6272,0.8975); rgb(94pt)=(0.1354,0.6317,0.8953); rgb(95pt)=(0.1321,0.6363,0.8932); rgb(96pt)=(0.1288,0.6408,0.891); rgb(97pt)=(0.1253,0.6453,0.8887); rgb(98pt)=(0.1219,0.6497,0.8862); rgb(99pt)=(0.1185,0.6541,0.8834); rgb(100pt)=(0.1152,0.6584,0.8804); rgb(101pt)=(0.1119,0.6627,0.877); rgb(102pt)=(0.1085,0.6669,0.8734); rgb(103pt)=(0.1048,0.671,0.8695); rgb(104pt)=(0.1009,0.675,0.8653); rgb(105pt)=(0.0964,0.6789,0.8609); rgb(106pt)=(0.0914,0.6828,0.8562); rgb(107pt)=(0.0855,0.6865,0.8513); rgb(108pt)=(0.0789,0.6902,0.8462); rgb(109pt)=(0.0713,0.6938,0.8409); rgb(110pt)=(0.0628,0.6972,0.8355); rgb(111pt)=(0.0535,0.7006,0.8299); rgb(112pt)=(0.0433,0.7039,0.8242); rgb(113pt)=(0.0328,0.7071,0.8183); rgb(114pt)=(0.0234,0.7103,0.8124); rgb(115pt)=(0.0155,0.7133,0.8064); rgb(116pt)=(0.0091,0.7163,0.8003); rgb(117pt)=(0.0046,0.7192,0.7941); rgb(118pt)=(0.0019,0.722,0.7878); rgb(119pt)=(0.0009,0.7248,0.7815); rgb(120pt)=(0.0018,0.7275,0.7752); rgb(121pt)=(0.0046,0.7301,0.7688); rgb(122pt)=(0.0094,0.7327,0.7623); rgb(123pt)=(0.0162,0.7352,0.7558); rgb(124pt)=(0.0253,0.7376,0.7492); rgb(125pt)=(0.0369,0.74,0.7426); rgb(126pt)=(0.0504,0.7423,0.7359); rgb(127pt)=(0.0638,0.7446,0.7292); rgb(128pt)=(0.077,0.7468,0.7224); rgb(129pt)=(0.0899,0.7489,0.7156); rgb(130pt)=(0.1023,0.751,0.7088); rgb(131pt)=(0.1141,0.7531,0.7019); rgb(132pt)=(0.1252,0.7552,0.695); rgb(133pt)=(0.1354,0.7572,0.6881); rgb(134pt)=(0.1448,0.7593,0.6812); rgb(135pt)=(0.1532,0.7614,0.6741); rgb(136pt)=(0.1609,0.7635,0.6671); rgb(137pt)=(0.1678,0.7656,0.6599); rgb(138pt)=(0.1741,0.7678,0.6527); rgb(139pt)=(0.1799,0.7699,0.6454); rgb(140pt)=(0.1853,0.7721,0.6379); rgb(141pt)=(0.1905,0.7743,0.6303); rgb(142pt)=(0.1954,0.7765,0.6225); rgb(143pt)=(0.2003,0.7787,0.6146); rgb(144pt)=(0.2061,0.7808,0.6065); rgb(145pt)=(0.2118,0.7828,0.5983); rgb(146pt)=(0.2178,0.7849,0.5899); rgb(147pt)=(0.2244,0.7869,0.5813); rgb(148pt)=(0.2318,0.7887,0.5725); rgb(149pt)=(0.2401,0.7905,0.5636); rgb(150pt)=(0.2491,0.7922,0.5546); rgb(151pt)=(0.2589,0.7937,0.5454); rgb(152pt)=(0.2695,0.7951,0.536); rgb(153pt)=(0.2809,0.7964,0.5266); rgb(154pt)=(0.2929,0.7975,0.517); rgb(155pt)=(0.3052,0.7985,0.5074); rgb(156pt)=(0.3176,0.7994,0.4975); rgb(157pt)=(0.3301,0.8002,0.4876); rgb(158pt)=(0.3424,0.8009,0.4774); rgb(159pt)=(0.3548,0.8016,0.4669); rgb(160pt)=(0.3671,0.8021,0.4563); rgb(161pt)=(0.3795,0.8026,0.4454); rgb(162pt)=(0.3921,0.8029,0.4344); rgb(163pt)=(0.405,0.8031,0.4233); rgb(164pt)=(0.4184,0.803,0.4122); rgb(165pt)=(0.4322,0.8028,0.4013); rgb(166pt)=(0.4463,0.8024,0.3904); rgb(167pt)=(0.4608,0.8018,0.3797); rgb(168pt)=(0.4753,0.8011,0.3691); rgb(169pt)=(0.4899,0.8002,0.3586); rgb(170pt)=(0.5044,0.7993,0.348); rgb(171pt)=(0.5187,0.7982,0.3374); rgb(172pt)=(0.5329,0.797,0.3267); rgb(173pt)=(0.547,0.7957,0.3159); rgb(175pt)=(0.5748,0.7929,0.2941); rgb(176pt)=(0.5886,0.7913,0.2833); rgb(177pt)=(0.6024,0.7896,0.2726); rgb(178pt)=(0.6161,0.7878,0.2622); rgb(179pt)=(0.6297,0.7859,0.2521); rgb(180pt)=(0.6433,0.7839,0.2423); rgb(181pt)=(0.6567,0.7818,0.2329); rgb(182pt)=(0.6701,0.7796,0.2239); rgb(183pt)=(0.6833,0.7773,0.2155); rgb(184pt)=(0.6963,0.775,0.2075); rgb(185pt)=(0.7091,0.7727,0.1998); rgb(186pt)=(0.7218,0.7703,0.1924); rgb(187pt)=(0.7344,0.7679,0.1852); rgb(188pt)=(0.7468,0.7654,0.1782); rgb(189pt)=(0.759,0.7629,0.1717); rgb(190pt)=(0.771,0.7604,0.1658); rgb(191pt)=(0.7829,0.7579,0.1608); rgb(192pt)=(0.7945,0.7554,0.157); rgb(193pt)=(0.806,0.7529,0.1546); rgb(194pt)=(0.8172,0.7505,0.1535); rgb(195pt)=(0.8281,0.7481,0.1536); rgb(196pt)=(0.8389,0.7457,0.1546); rgb(197pt)=(0.8495,0.7435,0.1564); rgb(198pt)=(0.86,0.7413,0.1587); rgb(199pt)=(0.8703,0.7392,0.1615); rgb(200pt)=(0.8804,0.7372,0.165); rgb(201pt)=(0.8903,0.7353,0.1695); rgb(202pt)=(0.9,0.7336,0.1749); rgb(203pt)=(0.9093,0.7321,0.1815); rgb(204pt)=(0.9184,0.7308,0.189); rgb(205pt)=(0.9272,0.7298,0.1973); rgb(206pt)=(0.9357,0.729,0.2061); rgb(207pt)=(0.944,0.7285,0.2151); rgb(208pt)=(0.9523,0.7284,0.2237); rgb(209pt)=(0.9606,0.7285,0.2312); rgb(210pt)=(0.9689,0.7292,0.2373); rgb(211pt)=(0.977,0.7304,0.2418); rgb(212pt)=(0.9842,0.733,0.2446); rgb(213pt)=(0.99,0.7365,0.2429); rgb(214pt)=(0.9946,0.7407,0.2394); rgb(215pt)=(0.9966,0.7458,0.2351); rgb(216pt)=(0.9971,0.7513,0.2309); rgb(217pt)=(0.9972,0.7569,0.2267); rgb(218pt)=(0.9971,0.7626,0.2224); rgb(219pt)=(0.9969,0.7683,0.2181); rgb(220pt)=(0.9966,0.774,0.2138); rgb(221pt)=(0.9962,0.7798,0.2095); rgb(222pt)=(0.9957,0.7856,0.2053); rgb(223pt)=(0.9949,0.7915,0.2012); rgb(224pt)=(0.9938,0.7974,0.1974); rgb(225pt)=(0.9923,0.8034,0.1939); rgb(226pt)=(0.9906,0.8095,0.1906); rgb(227pt)=(0.9885,0.8156,0.1875); rgb(228pt)=(0.9861,0.8218,0.1846); rgb(229pt)=(0.9835,0.828,0.1817); rgb(230pt)=(0.9807,0.8342,0.1787); rgb(231pt)=(0.9778,0.8404,0.1757); rgb(232pt)=(0.9748,0.8467,0.1726); rgb(233pt)=(0.972,0.8529,0.1695); rgb(234pt)=(0.9694,0.8591,0.1665); rgb(235pt)=(0.9671,0.8654,0.1636); rgb(236pt)=(0.9651,0.8716,0.1608); rgb(237pt)=(0.9634,0.8778,0.1582); rgb(238pt)=(0.9619,0.884,0.1557); rgb(239pt)=(0.9608,0.8902,0.1532); rgb(240pt)=(0.9601,0.8963,0.1507); rgb(241pt)=(0.9596,0.9023,0.148); rgb(242pt)=(0.9595,0.9084,0.145); rgb(243pt)=(0.9597,0.9143,0.1418); rgb(244pt)=(0.9601,0.9203,0.1382); rgb(245pt)=(0.9608,0.9262,0.1344); rgb(246pt)=(0.9618,0.932,0.1304); rgb(247pt)=(0.9629,0.9379,0.1261); rgb(248pt)=(0.9642,0.9437,0.1216); rgb(249pt)=(0.9657,0.9494,0.1168); rgb(250pt)=(0.9674,0.9552,0.1116); rgb(251pt)=(0.9692,0.9609,0.1061); rgb(252pt)=(0.9711,0.9667,0.1001); rgb(253pt)=(0.973,0.9724,0.0938); rgb(254pt)=(0.9749,0.9782,0.0872); rgb(255pt)=(0.9769,0.9839,0.0805)}
]
\addplot [forget plot] graphics [xmin=-0.50125313283208, xmax=400.501253132832, ymin=0.491025641025641, ymax=1.20897435897436] {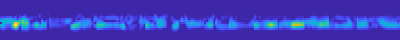};

\draw [red, line width=1.0mm, dashed, dash pattern=on 12pt off 12pt](axis cs:0,0.755) -- (axis cs:400,0.755);
\end{axis}

\end{tikzpicture}%
}}
	\caption{Range profile heat map showing a stationary target}
	\label{fig:heatmap}
	\vspace{-0.4 cm}
\end{figure}

 The target of interest is tracked to be stationary within the same range bin during the monitoring period, as shown in Fig.~\ref{fig:heatmap}. The phase of the estimated complex range bin is estimated and unwrapped, as shown in Fig.~\ref{fig:newcodebd}. The output of the bandpass filter reflects the induced micro motions in the subject of interest as shown in Fig.~\ref{fig:peaks}, where each peak is reflecting a moving harmonic, in which one of them is reflecting the respiratory rate. Due to monitoring in standing position, the filtered waveform will include harmonics reflecting the body balancing motion and the breathing activity. The main focus of the proposed algorithm is analyzing the captured harmonics for accurate estimation of the respiratory rate. Three different approaches are discussed briefly in the following sections.

\subsection{Spectral Estimation} \label{sec:spect}
\par The measurement protocol within the previous studies was held on sitting position, where a single and maximum harmonic is expected to reflect the true respiratory rate as shown in Fig.~\ref{fig:1peak}. Unlikely, for the standing position, the highest peak in the bandpass filter output is not always induced by the target's breathing activity but rather by the body balancing motion. Accordingly, depending only on the highest peak will not give the correct respiratory rate estimation. Instead, the first three highest peaks were extracted, and a confidence metric (CM) was calculated for each peak. The spectral estimation approach is implemented by first calculating the confidence metric then applying the extra step of adaptive averaging for enhancing the harmonic selection.     

\subsubsection{Confidence Metric} 
\par The CM is calculated as the ratio of the signal power of the peak and five frequency bins before and five frequency bins after the peak over the remaining frequency bins in the bandpass spectrum as shown in Equation~\ref{eq:cm}. The bigger the value of the CM, the more confident we use it in the respiratory rate estimation.  
\begin{equation}
    {Confidence Metric}_i = \frac{\sum{Peak_i}}{\sum{Signal} - \sum{Peak_i}}
\label{eq:cm}
\end{equation}

\subsubsection{Adaptive Averaging Techniques}
\label{sec:AdaptiveAveraging}
\par In order to efficiently estimate the respiratory rate in such a challenging setup as standing position, three adaptive averaging techniques have been introduced as follows:

\paragraph{CM Weighted} the overall estimated respiratory rate is calculated according to the respiratory rate of the highest three peaks. The summation of the three peaks multiplied by their corresponding CM value is divided by the total CM of the three peaks. The final value is used as the estimated respiratory rate as follows:
\begin{equation}
CM_{weighted} = \frac{\sum_{i=0}^{2}RR_i * CM_i}{\sum_{i=0}^{2}CM_i}
\label{eq:cmweighted}
\end{equation}
        
\paragraph{Power Weighted} the overall output of the bandpass filter is initially converted to respiratory rates. Then the summation of the respiratory rates in the signal is multiplied by its corresponding power value and divided by the total power. The final value is used as the estimated respiratory rate, as shown in the following equation:
        
        \begin{equation}
            Power_{weighted} = \frac{\sum_{i=0}^{SignalLen}RR_i * Power_i}{\sum_{i=0}^{SignalLen}Power_i}
        \label{eq:powerweighted}
        \end{equation}

        \paragraph{Power Peaks Weighted} the overall estimated respiratory rate is calculated according to the respiratory rate of the highest three peaks. The summation of the three peaks multiplied by their corresponding power value is divided by the total power of the three peaks. The final value is used as the estimated respiratory rate, as shown in the following equation:
        
        \begin{equation}
            PowerPeak_{weighted} = \frac{\sum_{i=0}^{2}RR_i * Power_i}{\sum_{i=0}^{2}Power_i}
        \label{eq:powerpeakweighted}
        \end{equation}

\subsection{Deep Learning Approach}
\label{sec:DeepLearning}
\begin{figure*}[t!]
     \centering
     \begin{subfigure}[ht]{0.45\textwidth}
         \centering
            {\resizebox{0.98\columnwidth}{!}{
%
%
\definecolor{mycolor1}{rgb}{0.00000,0.44700,0.74100}%
\definecolor{mycolor2}{rgb}{0.85000,0.32500,0.09800}%
\definecolor{mycolor3}{rgb}{0.92900,0.69400,0.12500}%
\definecolor{mycolor4}{rgb}{0.49400,0.18400,0.55600}%
\definecolor{mycolor5}{rgb}{0.46600,0.67400,0.18800}%
\begin{tikzpicture}

\begin{axis}[%
width=4.521in,
height=2.5in,
at={(2.5in,1.608in)},
scale only axis,
clip=false,
xmin=0,
xmax=50,
xtick distance=10,
ytick distance=10000000000000,
xlabel style={font=\color{white!15!black}},
xlabel={\Large Respiration Rate [BPM]},
ytick style={draw=none},
xtick style={draw=none},
ymin=0,
ymax=34512554967040,
ylabel style={font=\color{white!15!black}},
ylabel={\Large Power [A.U.]},
axis background/.style={fill=white},
legend style={legend cell align=left, align=left, draw=white!15!black}
]
\addplot [color=mycolor1, line width=2.0pt, forget plot]
  table[row sep=crcr]{%
0	596143046656\\
1.171875	830611259392\\
2.34375	1312313311232\\
3.515625	2493884923904\\
4.6875	6432209502208\\
5.859375	14548177256448\\
7.03125	24418319335424\\
8.203125	30512554967040\\
9.375	29112842321920\\
10.546875	21970494160896\\
11.71875	14630420217856\\
12.890625	11589724930048\\
14.0625	13320573681664\\
15.234375	17125426069504\\
16.40625	19991084662784\\
17.578125	20822039199744\\
18.75	20541320724480\\
19.921875	20110743961600\\
21.09375	18589604118528\\
22.265625	14076793061376\\
23.4375	7129006080000\\
24.609375	2054067453952\\
25.78125	3061769306112\\
26.953125	9182549901312\\
28.125	14592173408256\\
29.296875	14509873823744\\
30.46875	9629771759616\\
31.640625	4228598202368\\
32.8125	1278334468096\\
33.984375	564020641792\\
35.15625	570754859008\\
36.328125	468456374272\\
37.5	222731452416\\
38.671875	49043931136\\
39.84375	210810781696\\
41.015625	727887708160\\
42.1875	1099941543936\\
43.359375	829463527424\\
44.53125	222672093184\\
45.703125	63669817344\\
};
\addplot [color=mycolor2, line width=2.0pt]
  table[row sep=crcr]{%
14	0\\
14	30512554967040\\
};
\addlegendentry{Target}

\addplot [color=mycolor3, line width=2.0pt]
  table[row sep=crcr]{%
16.2097506838281	0\\
16.2097506838281	30512554967040\\
};
\addlegendentry{Power Weighted}

\addplot [color=mycolor4, line width=2.0pt]
  table[row sep=crcr]{%
16.6583414296157	0\\
16.6583414296157	30512554967040\\
};
\addlegendentry{CM Weighted}

\addplot [color=mycolor5, line width=2.0pt]
  table[row sep=crcr]{%
16.0103246208785	0\\
16.0103246208785	30512554967040\\
};
\addlegendentry{Power Peaks Weighted}

\addplot[only marks, mark=*, mark options={}, mark size=1.5000pt, color=red, fill=red, forget plot] table[row sep=crcr]{%
x	y\\
8.203125	30512554967040\\
17.578125	20822039199744\\
28.125	14592173408256\\
42.1875	1099941543936\\
};
\node[right, align=left,yshift=4ex,xshift=2ex]
at (axis cs:-0.001,29162891322080.797) {CM = 2.5836};
\node[right, align=left]
at (axis cs:17.083,22182253148073.355) {CM = 10.3253};
\node[right, align=left]
at (axis cs:26.316,15726193338632.584) {CM = 0.94183};
\node[right, align=left]
at (axis cs:39.358,2288970643682.267) {CM = 0.06066};
\end{axis}
\end{tikzpicture}%
}}    	
            \caption{Three main harmonics reflecting the body micro motions}
    	\label{fig:3peaks}
     \end{subfigure}
     \begin{subfigure}[ht]{0.45\textwidth}
         \centering
         {\resizebox{0.98\columnwidth}{!}{
%
%
\definecolor{mycolor1}{rgb}{0.00000,0.44700,0.74100}%
\definecolor{mycolor2}{rgb}{0.85000,0.32500,0.09800}%
\definecolor{mycolor3}{rgb}{0.92900,0.69400,0.12500}%
\definecolor{mycolor4}{rgb}{0.49400,0.18400,0.55600}%
\definecolor{mycolor5}{rgb}{0.46600,0.67400,0.18800}%
\begin{tikzpicture}

\begin{axis}[%
width=4.521in,
height=2.5in,
at={(2.08in,1.105in)},
scale only axis,
clip=false,
xmin=0,
xmax=50,
xtick distance=10,
ytick distance=10000000000000,
xlabel style={font=\color{white!15!black}},
xlabel={\Large Respiration Rate [BPM]},
ymin=0,
ymax=34512554967040,
ylabel style={font=\color{white!15!black}},
ylabel={\Large Power [A.U.]},
axis background/.style={fill=white},
legend style={legend cell align=left, align=left, draw=white!15!black}
]
\addplot [color=mycolor1, line width=2.0pt, forget plot]
  table[row sep=crcr]{%
0	729111068672\\
1.171875	838180012032\\
2.34375	1185767096320\\
3.515625	1785111248896\\
4.6875	2521850707968\\
5.859375	3077650251776\\
7.03125	3046137921536\\
8.203125	2265426165760\\
9.375	1103385591808\\
10.546875	305770004480\\
11.71875	346164494336\\
12.890625	848624680960\\
14.0625	920767823872\\
15.234375	459189977088\\
16.40625	1128180744192\\
17.578125	5400435884032\\
18.75	13867801378816\\
19.921875	23445381316608\\
21.09375	28965970378752\\
22.265625	27285428109312\\
23.4375	19938666348544\\
24.609375	11594097491968\\
25.78125	6022994329600\\
26.953125	3737898713088\\
28.125	3079021527040\\
29.296875	2597127716864\\
30.46875	1963119476736\\
31.640625	1295767830528\\
32.8125	622516109312\\
33.984375	114097061888\\
35.15625	158433411072\\
36.328125	797291839488\\
37.5	1422882897920\\
38.671875	1425529765888\\
39.84375	977502601216\\
41.015625	794848788480\\
42.1875	1164323717120\\
43.359375	1586389450752\\
44.53125	1498264502272\\
45.703125	1038838202368\\
};
\addplot [color=mycolor2, line width=2.0pt]
  table[row sep=crcr]{%
20	0\\
20	28965970378752\\
};
\addlegendentry{Target}

\addplot [color=mycolor3, line width=2.0pt]
  table[row sep=crcr]{%
21.673418363609	0\\
21.673418363609	28965970378752\\
};
\addlegendentry{Power Weighted}

\addplot [color=mycolor4, line width=2.0pt]
  table[row sep=crcr]{%
18.0496460565643	0\\
18.0496460565643	28965970378752\\
};
\addlegendentry{CM Weighted}

\addplot [color=mycolor5, line width=2.0pt]
  table[row sep=crcr]{%
22.7322954148422	0\\
22.7322954148422	28965970378752\\
};
\addlegendentry{Power Peaks Weighted}

\addplot[only marks, mark=*, mark options={}, mark size=1.5000pt, color=red, fill=red, forget plot] table[row sep=crcr]{%
x	y\\
21.09375	28965970378752\\
43.359375	1586389450752\\
38.671875	1425529765888\\
14.0625	920767823872\\
};
\node[right, align=left,yshift=4ex,xshift=-6ex]
at (axis cs:23.068,28389304735570.184) {CM = 5.4523};
\node[right, align=left,yshift=1ex,xshift=-1ex]
at (axis cs:39.453,4622541081479.618) {CM = 0.088355};
\node[right, align=left,yshift=1ex]
at (axis cs:32.407,2401039931603.839) {CM = 0.16718};
\node[right, align=left,yshift=1ex,xshift=-2ex]
at (axis cs:8.584,2685438679379.015) {CM = 5.5883};
\end{axis}

\end{tikzpicture}%
}}
    	\caption{Single harmonic expected to be reflecting the chest activity}
    	\label{fig:1peak}
     \end{subfigure}
        \caption{Bandpass filter output signal waveform}
        \label{fig:peaks}
        \vspace{-0.5 cm}
\end{figure*}
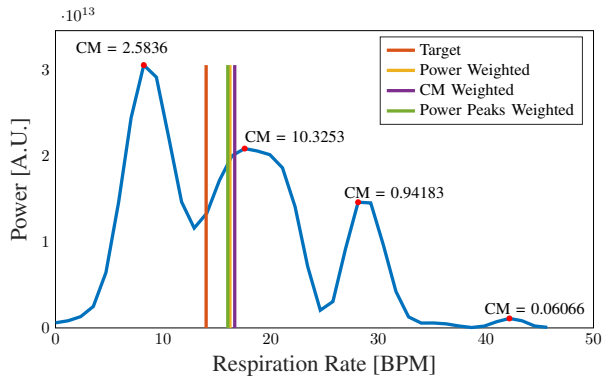
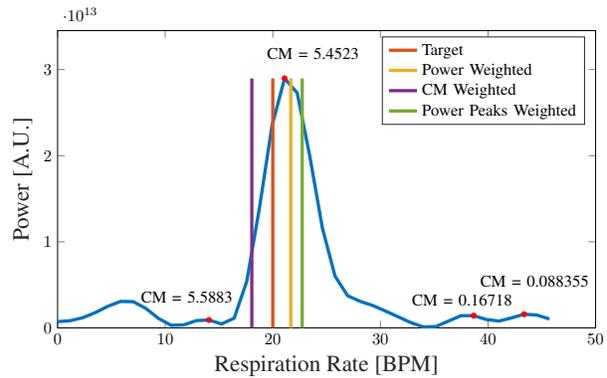
\par The aforementioned approaches didn't show a reliable performance due to the unstable harmonic behavior. Accordingly, a neural-network-based regressor is implemented, taking the three averaging techniques as inputs, in addition to the respiration rate, power values, and CM of the highest three peaks. The output of the model gives a final respiration rate prediction. The proposed network consists of five neural network layers. The first four use Rectified Linear Unit (ReLU) as a non-linear activation function, followed by a regression layer at the end. A trade-off between the proposed model complexity and the efficient prediction experimented. The model is integrated with the real-time processing chain, and the previously mentioned configuration was found to be the most accurate in the prediction yet meeting the real-time processing requirement.

\par The proposed model showed accurate estimation in the mid-range of the respiratory rate values(\unit[10-25]{bpm}). However, the error rate was found to be higher at the extremities of the respiratory rate range. This is due to the difficulty of collecting reliable datasets with very high or low respiration rates to be used for the learning process. A brief discussion is presented in Sec.~\ref{sec:results}

\subsection{Adaptive Peak Selection}
\label{sec:AdaptivePeak}
\par In order to make benefit from the CM concept and to overcome the learning technique limitations, a final adaptive peak selection technique integrated with a Kalman filter~\cite{kalman1960new} is proposed for more stabilized, accurate, and time-efficient respiratory rate estimation. In this final integration shown in Fig.~\ref{fig:specest}, we extract the higher four peaks in our breathing waveform. The CM is computed for them. An example can be seen in Fig.~\ref{fig:3peaks} where the CM is calculated for the maximum four peaks. The peak with the maximum CM is chosen as the measurement of the state for the Kalman filter. 

\par In order to implement a Kalman filter in our signal processing chain, we have to think about the sensor noise covariance matrix ($R$) because this parameter directly controls the value of the Kalman gain. The CM is used to estimate $R$. The algorithm used to calculate the $R$ value is as follows: The peaks with the highest two confidence metrics are found, and the ratio of each one with respect to the total power of the four peaks is calculated. If one of them has a ratio higher than 0.5, the value of the highest confidence metric is used to calculate the $R$ value. Otherwise, a weighted average is taken between the highest two confidence metrics. The output from the Kalman filter is the final respiratory rate estimation.

\begin{table}[b]
\centering
\caption{Vernier Go Direct\textsuperscript{\textregistered} Respiration Belt specifications}
\label{tab:rbspecs}
\resizebox{\columnwidth}{!}{%
\begin{tabular}{@{}lllllllc@{}}
\toprule
Parameter                                                                &    &  &  &  &  & Value  \\ \midrule
Range                                                                    &    &  &  &  &  & 0-50 N \\
Resolution                                                               &    &  &  &  &  & 0.01 N \\
Response Time                                                            &   &  &  &  &  & 50 ms  \\
\begin{tabular}[c]{@{}c@{}}Respiration Rate Sample Window\end{tabular} &  &    &  &  &  & 10 s   \\ \bottomrule
\end{tabular}%
}
\end{table}


\begin{figure*}[ht]
     \centering
     \begin{subfigure}[ht]{0.45\textwidth}
         \centering
            {\resizebox{0.98\columnwidth}{!}{
%
%
\definecolor{mycolor1}{rgb}{0.00000,0.44700,0.74100}%
\definecolor{mycolor2}{rgb}{0.85000,0.32500,0.09800}%
\begin{tikzpicture}

\begin{axis}[%
width=4.521in,
height=2.5in,
at={(2.08in,1.182in)},
scale only axis,
xmin=0,
xmax=10,
xlabel style={font=\color{white!15!black}},
xlabel={\Large Mean Absolute Error},
ymin=0,
ymax=14,
ylabel style={font=\color{white!15!black}},
ylabel={\Large Target Count},
axis background/.style={fill=white},
axis x line*=bottom,
axis y line*=left,
legend style={legend cell align=left, align=left, fill=none, draw=none}
]
\addplot[ybar interval, fill=mycolor1, fill opacity=0.6, draw=black, area legend] table[row sep=crcr] {%
x	y\\
1.2	2\\
1.85	5\\
2.5	8\\
3.15	14\\
3.8	5\\
4.45	4\\
5.1	2\\
5.75	1\\
6.4	1\\
};
\addlegendentry{Classical Technique}

\addplot[ybar interval, fill=mycolor2, fill opacity=0.6, draw=black, area legend] table[row sep=crcr] {%
x	y\\
0	4\\
0.7	8\\
1.4	12\\
2.1	6\\
2.8	6\\
3.5	3\\
4.2	2\\
4.9	2\\
};
\addlegendentry{Proposed Technique}

\end{axis}

\end{tikzpicture}%
}}    	
            \caption{\centering Mean Absolute Error with respect to the ground truth of the medical device}
    	\label{fig:measuresub6}
     \end{subfigure}
     \begin{subfigure}[ht]{0.45\textwidth}
         \centering
         {\resizebox{0.98\columnwidth}{!}{
%
%
\definecolor{mycolor1}{rgb}{0.00000,0.44700,0.74100}%
\definecolor{mycolor2}{rgb}{0.85000,0.32500,0.09800}%
\begin{tikzpicture}

\begin{axis}[%
width=4.521in,
height=2.5in,
at={(2.08in,1.182in)},
scale only axis,
xmin=0,
xmax=10,
xlabel style={font=\color{white!15!black}},
xlabel={\Large Standard Deviation},
ymin=0,
ymax=16,
ylabel style={font=\color{white!15!black}},
ylabel={\Large Target Count},
axis background/.style={fill=white},
axis x line*=bottom,
axis y line*=left,
legend style={legend cell align=left, align=left, fill=none, draw=none}
]
\addplot[ybar interval, fill=mycolor1, fill opacity=0.6, draw=black, area legend] table[row sep=crcr] {%
x	y\\
0.6	2\\
1.26	2\\
1.92	2\\
2.58	12\\
3.24	10\\
3.9	6\\
4.56	6\\
5.22	1\\
5.88	1\\
};
\addlegendentry{Classical Technique}

\addplot[ybar interval, fill=mycolor2, fill opacity=0.6, draw=black, area legend] table[row sep=crcr] {%
x	y\\
0	5\\
0.32	16\\
0.64	9\\
0.96	5\\
1.28	5\\
1.6	1\\
1.92	1\\
};
\addlegendentry{Proposed Technique}

\end{axis}

\end{tikzpicture}%
}}
    	\caption{\centering Standard Deviation of respiratory rate reflecting the stability of measurements due to the Kalman filter}
    	\label{fig:measuresub1}
     \end{subfigure}
        \caption{Error analysis for comparing both the classical and adaptive peak selection approaches}
        \label{fig:results}
        \vspace{-0.2 cm}
\end{figure*}
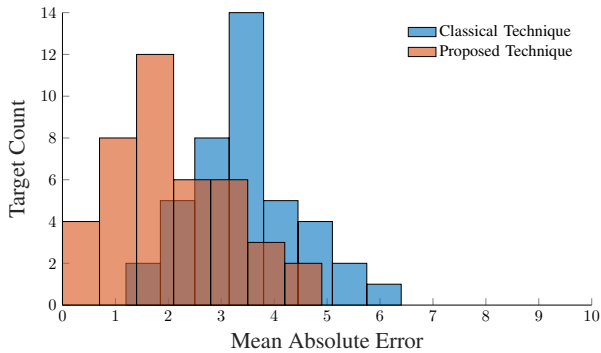
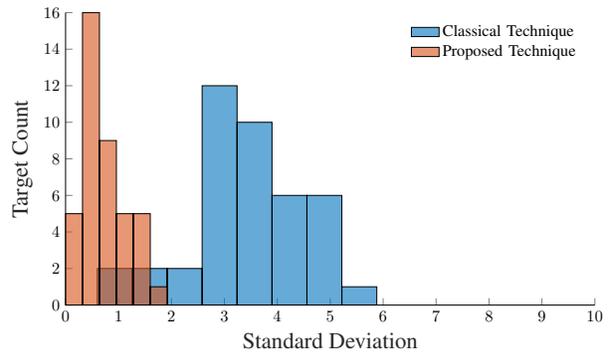

\section{Experimental Setup}\label{sec:exp}

\par
\begin{figure}[t]
    \centering
    \includegraphics[width=0.35\textwidth]{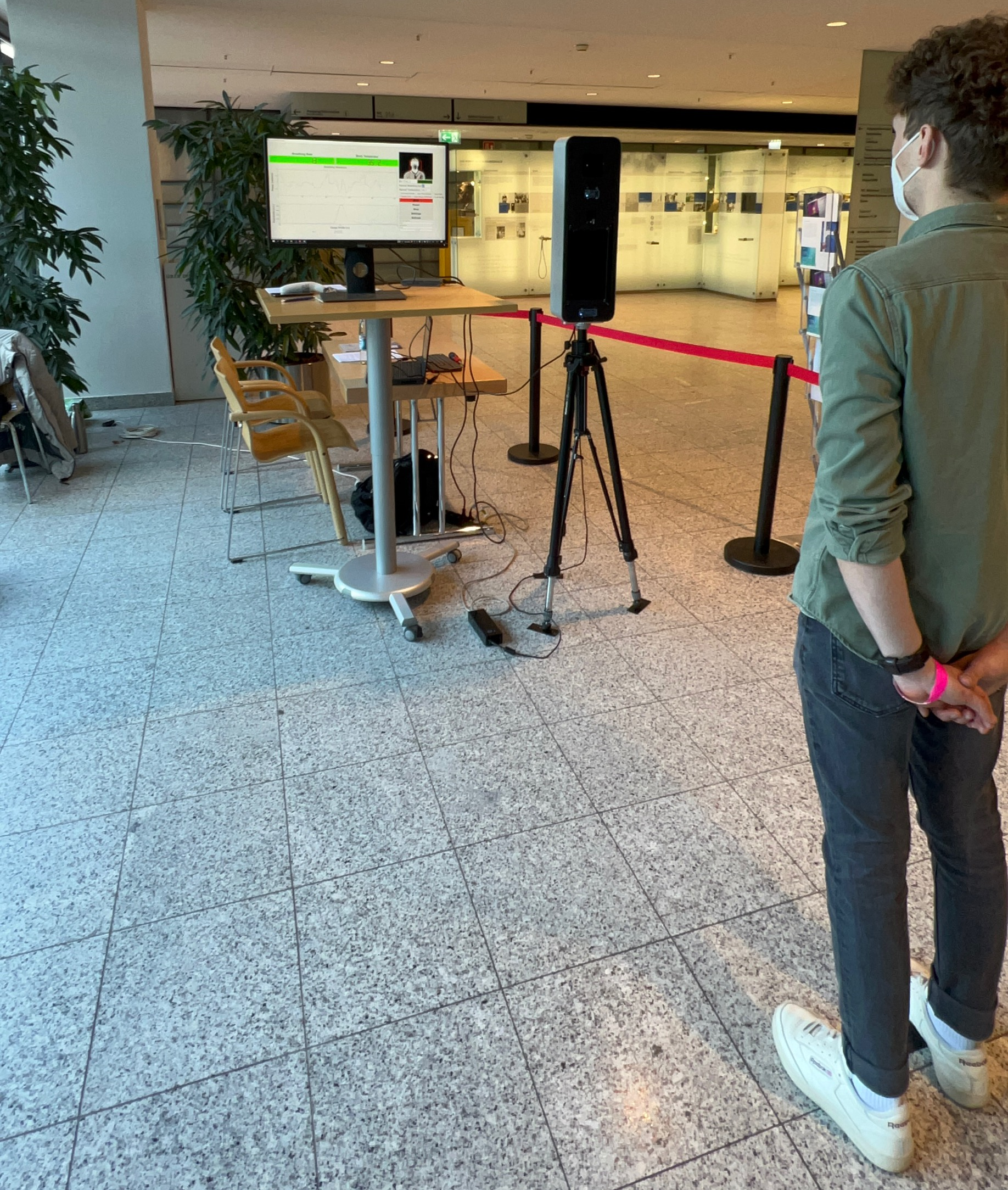}
	\caption{Test target example in the public hospital entrance}
	\label{fig:real-setup}
	\vspace{-0.3 cm}
\end{figure}

 The radar is parametrized according to Table~\ref{tab:radarspecs}, was mounted on a height of about \unit[1.3]{m} to be facing the abdominal area of the human body for heights of \unit[1.5-1.9]{m} not necessarily the chest. The radar is integrated with a thermal camera that is used for face detection for estimating the body temperature. The overall system can be overlooked here \cite{fraunhofer_2020}. However, our study is mainly focused on the radar algorithm for respiratory rate monitoring. The targets were asked to be standing at a distance of \unit[1.0-1.5]{m} from the radar as shown in  Fig.~\ref{fig:real-setup}. The initial experiments were conducted in the labs of the Fraunhofer facility for comparing the three main proposed approaches, where the adaptive peak selection approach has shown the best performance. A discussion about the results of each technique is presented in Sec.~\ref{sec:results}.

\par A feasibility study was held in the Robert-Bosch-hospital in Stuttgart, Germany, where \unit[41]{targets} have shared in the study. The targets were with a height range of \unit[1.5-1.9]{m}, age range of \unit[22-72]{years old} and different fitness levels. A Vernier Go Direct\textsuperscript{\textregistered} Respiration Belt \cite{vernier} with specification in Table~\ref{tab:rbspecs} was used as the ground truth for our measurements. For each subject, the radar was used to capture 500 samples for both the classical technique and our proposed technique. The algorithm is implemented on the micro-controller of the proposed radar chip and could achieve a sampling rate of $\approx$\unit[17]{Hz}. Each sample here includes an estimation from the classical technique and the adaptive peak selection approach since it has shown the most reliable performance in the lab tests.

\begin{table}[b]
\centering
\caption{MIMO radar module parametrization.}
\label{tab:radarspecs}
\resizebox{\columnwidth}{!}{%
\begin{tabular}{@{}llllllll@{}}
\toprule
\multicolumn{1}{c}{Radar Parametrization} &  &  &  &  &  &  & Value              \\ \midrule
Carrier Frequency ($f_0$)                 &  &  &  &  &  &  & \unit[60-64]{GHz}  \\
Tx-Rx antennas                            &  &  &  &  &  &  & 3-4                \\
Bandwidth ($B$)                           &  &  &  &  &  &  & \unit[3.7]{GHz}    \\
Chirp duration ($T_c$)                    &  &  &  &  &  &  & \unit[57]{$\mu s$} \\
Samples per chirp ($N_S$)                 &  &  &  &  &  &  & 200                \\
Chirps per frame ($N_P$)                  &  &  &  &  &  &  & 2                  \\ \bottomrule
\end{tabular}%
}
\end{table}


\section{Results and Discussion}\label{sec:results}

\par Holding the experimental setup on the standing position has been proven to be more challenging than sitting or sleeping while stationary. The output signal waveform from the bandpass filter includes multiple harmonics induced due to the body micro-motion for body balancing. Therefore, unlike the state-of-the-art techniques, depending only on the highest peak of the signal will not result in an accurate estimation of the respiratory rate. Accordingly, as described in Sec.~\ref{sec:AdaptiveAveraging}, the three adaptive averaging techniques were used to get a more accurate estimation for the respiratory rate, utilizing the data of the three peaks with the highest CM value. However, none of the adaptive averaging techniques was reliable enough or showed enough robustness against the highly-changing bandpass signal waveform. 

\par In order to get a final result including all the averaging techniques, a neural-network-based regressor was implemented as described in Sec.~\ref{sec:DeepLearning}. The learning approach showed a very good performance on the mid-range respiratory rate values. However, it showed poor performance at the extremities of the respiratory rate range. Table~\ref{tab:DeepLearning} shows the deviation of the error values with respect to the ground truth. Also, increasing the model complexity was not possible due to the limited learning dataset and the real-time operation requirements. Finally, to overcome the learning approach limitations, the adaptive peak selection technique mentioned in Sec.~\ref{sec:AdaptivePeak} is used. 

\par As mentioned in Sec.~\ref{sec:exp}, the adaptive peak selection algorithm has been tested in a feasibility study that is conducted on \unit[41]{subjects}. As shown in Fig.~\ref{fig:measuresub6}, the adaptive peak selection has shown an overall better performance compared to the classical technique in which a less absolute error of \unit[$\approx 1.5$]{bpm} is achieved. Additionally, adding the Kalman filter has enhanced the estimation stability, as shown in Fig.~\ref{fig:measuresub1}. Each measurement for each subject shows less varying estimations for the respiratory rate.  

\begin{table}[b]
\caption{Deep Learning Approach Error Rates}
\centering
\resizebox{\columnwidth}{!}{%
\begin{tabular}{@{}llccc@{}}
\toprule
  & BR Category & \multicolumn{1}{l}{Manual BR Range} & &\multicolumn{1}{l}{Error Rate} \\ \midrule
1 & Low         & 6 - 14              &                & 7.97   \\
2 & Mid         & 15 - 24             &                & 2.16   \\
3 & High        & 25 - 36              &               & 9.32   \\ 
\bottomrule
\end{tabular}
\label{tab:DeepLearning}
}
\end{table}
\section{Conclusion}\label{sec:conc}

 \par The non-visionary capturing capability of the radar sensors presents it as a feasible sensor for our application. We focused on respiratory rate measurement for standing people accurately on a contactless-basis. We developed an algorithm that consisted of several stages. Our system can be easily deployed in hospitals and airports to quickly estimate the respiratory rate of a person and provide a good indication about whether this person is infected with respiratory disease or not. The proposed algorithm, including the adaptive peak selection and the Kalman filter, has shown the best estimation performance. It was tested on 41 subjects in a feasibility study that was held in a public hospital. The targets were of different heights, ages, and fitness levels. The system has shown an error rate of $\approx\pm\unit[1.5]{bpm}$ compared to the Vernier Go Direct\textsuperscript{\textregistered} respiration belt. The Kalman filter has a direct effect on enhancing the estimation stability of the system. 

\par The results we acquired and presented still had a minimal margin of error that will be investigated as future work. The heart rate estimation will be investigated in the next phase. Some future work possibilities can be done to improve the sensitivity of the Kalman filter. Moreover, this work can be extended to include a monitoring capability for multiple target simultaneously.

\bibliographystyle{IEEEtran}

\end{document}